\documentclass[a4paper]{article}
\usepackage{array,amsfonts,amsmath,graphicx, xcolor, graphics}
\usepackage{INTERSPEECH2019}
\usepackage{libertine}
\usepackage{multirow,array}
\usepackage[font=small]{caption}
\usepackage{subcaption}

\pdfoutput=1

\title{Direct Modelling of Speech Emotion from Raw Speech}
\name{Siddique Latif$^{1,2}$, Rajib Rana$^1$, Sara Khalifa$^{2,3}$,  Raja Jurdak$^2$, Julien Epps$^3$}
\address{
  $^1$University of Southerns Queensland, Australia\\
  $^2$Distributed Sensing Systems Group, Data61, CSIRO Australia\\
  $^3$University of New South Wales, Australia}
\email{siddique.latif@usq.edu.au}

\begin{document}

\maketitle
\begin{abstract}
Speech emotion recognition is a challenging task and heavily depends on hand-engineered acoustic features, which are typically crafted to echo human perception of speech signals. However, a filter bank that is designed from perceptual evidence is not always guaranteed to be the best in a statistical modelling framework where the end goal is for example emotion classification. This has fuelled the emerging trend of learning representations from raw speech especially using deep learning neural networks. In particular, a combination of Convolution Neural Networks (CNNs) and Long Short Term Memory (LSTM) have gained great traction for the intrinsic property of LSTM in learning contextual information crucial for emotion recognition; and CNNs been used for its ability to overcome the scalability problem of regular neural networks. In this paper, we show that there are still opportunities to improve the performance of emotion recognition from the raw speech by exploiting the properties of CNN in modelling contextual information. We propose the use of parallel convolutional layers to harness multiple temporal resolutions in the feature extraction block that is jointly trained with the LSTM based classification network for the emotion recognition task. Our results suggest that the proposed model can reach the performance of CNN trained with hand-engineered features from both IEMOCAP and MSP-IMPROV datasets.

\end{abstract}

\section{Introduction}

Automatic speech emotion recognition has many important applications, such as diagnosis of depression \cite{zhu2018automated}, distress \cite{rana2019automated}, monitoring mood state
for bipolar patients \cite{gideon2016mood,rana2016context}, and so on. However, emotion recognition from speech is a complex task as emotional expression could vary significantly due to a multitude of contextual factors including culture, age, gender, accent, surrounding environment and so on~\cite{Rana16}. 

Research on speech emotion recognition primarily focuses on hand-engineered acoustic features as well as on designing efficient machine learning based models for accurate emotion prediction \cite{schuller2018speech,latif2018transfer}. In particular, building an appropriate feature representation and designing an appropriate classifier for these features have often been treated as separate problems in the speech recognition community. One drawback of this approach is that the designed features might not be best for the classification objective at hand. LogMel features have been the most popular feature to train Deep Neural Networks (DNNs) and their variants to date. The Mel filter bank is inspired by auditory and physiological evidence of how humans perceive speech signals~\cite{davis1980comparison}. However, based on the argument in~\cite{sainath2015learning} a filter bank that is designed from perceptual evidence is not always guaranteed to be the best filter bank in a statistical modelling framework where the end goal is emotion classification. These have led to a recent trend in the machine learning community towards deriving a representation of the input signal directly from raw, unprocessed data.  The network learns an intermediate representation of the raw input signal automatically that better suits the task at hand and hence lead to improved performance compared to the classical methods. 


A challenging issue in emotion recognition from speech is the efficient modelling of long temporal context \cite{tian2016recognizing}. This is because emotions are context-dependent \cite{Latif2018} and emotion specific information is embedded in the long temporal contexts \cite{sarma2018emotion}. LSTM \cite{hochreiter1997long} can model a long range of contexts due to the presence of a special structure called the memory cell.
This is why researchers frequently use LSTM for speech emotion recognition~\cite{tzirakis2018end}. Interestingly, convolutional layer filters can also be used to capture contextual information, which has shown great success in natural language processing (NLP) for sentence classification~\cite{zhang2015sensitivity}. In particular, multiple width filters can help improve performance because the model can simultaneously learn multiple contextual dependencies. This has also been validated for speech emotion recognition~\cite{aldeneh2017using}. However, these studies have generally used multiple width filters in a single layer. Recent studies have shown that parallel convolutional layers can extract  temporal information at multiple resolutions from the given data, which can improve the performance of the system \cite{yenigalla2018speech,zhang2018speech}. In contrast to using multiple filters in a single layer, in this paper we propose the use of parallel convolutional layers with different filter width to capture diverse contextual information from raw speech.  

The key contribution of this paper is the proposed network that consists of a multi-temporal CNN  stacked on  LSTM. The proposed construct of CNN provides an additional layer for capturing contextual information at mulitple temporal resolutions and is designed to complement LSTM for modelling long-term contextual information from raw speech.

\section{Related Work}

Many studies have considered the use of Deep Neural Networks (DNN) models for processing raw waveforms directly, but the majority of these are in the field of automatic speech recognition (ASR). Dimitri et al. \cite{palaz2015analysis} used a framework of CNN for ASR and achieved competitive results to standard short-term spectral features. The authors showed that convolution layers act like a data-driven filterbank and can model spectral envelope of raw speech. The authors in \cite{palaz2015convolutional}  showed that CNN can learn more generalised features across different
databases from raw speech compared to artificial neural networks and other feature-based approaches. The complementary approach of using CNN and LSTM jointly for raw speech has been evaluated in \cite{sainath2015learning,sainath2015convolutional}. The authors have shown that the use of LSTM with CNN, helps to reduce the word error rate and achieve competitive performance to the standard feature-based approaches. Besides ASR, researchers have highlighted the feature learning power of different DNN models from raw speech for many other tasks including environmental sound recognition~\cite{dai2017very}, speaker identification~\cite{muckenhirn2018towards,ravanelli2018speech}, and automatic tagging~\cite{dieleman2014end}.



Few studies have attempted to model emotions using raw speech with results not quite matching feature-based methods. For instance, two studies  \cite{trigeorgis2016adieu,tzirakis2018end} used end-to-end models by combining CNN and LSTM layers for predicting valence-activation on RECOLA database \cite{ringeval2013introducing} and achieved promising results. Sarma et al. \cite{sarma2018emotion} evaluated time-delay neural network (TDNN) based multiple architectures to model long term dependencies of speech emotion and provided promising results on IEMOCAP dataset. In~\cite{aldeneh2017using} multi-width filter CNN was applied to hand-engineered features (Mel Filterbanks (MFBs) which provided competitive results to the systems trained on popular emotional feature sets.  In contrast to previous studies \cite{sarma2018emotion,tzirakis2018end,trigeorgis2016adieu}, we propose a parallel configuration of convolutional layers with multiple filter lengths in feature extraction block to harness multiple temporal resolutions and simultaneously extract multiple contextual dependencies. The classification block is jointly optimised with the feature extraction block to achieve the emotion classification objective.



\section{Model}
\begin{figure}[!ht]
  \centering
  \includegraphics[trim=0.15cm 0.12cm 0.45cm 0.2cm,clip=true,width=\linewidth]{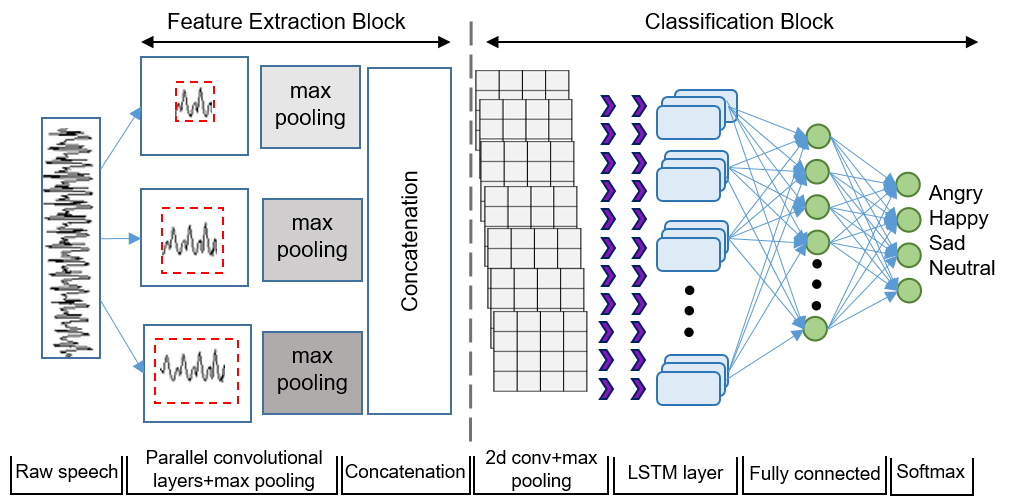}
  \caption{Schematic diagram of proposed model}
  \label{fig:model}
\end{figure}

Our model consists of two parts: a feature extraction block and a classification block, as shown in Figure \ref{fig:model}. 


\subsection{Feature extraction block}
\label{FE}

 In the feature extraction block, we use parallel convolutional layers with multiple filter lengths to capture both long-term and short-term interactions directly from raw speech. Given an input utterance, the convolutional layer identifies emotionally salient regions using finite impulse-response filters, since multiple filters with different lengths can capture diverse contextual dependencies simultaneously from the same region \cite{aldeneh2017using}. 

 In our model, $N$ parallel convolutional layers take $x_{t}$ as input of raw speech  and create $N$ different sequences of feature maps by convoluting $x_{t}$ with a set of filters of different lengths. The output of convoluting each layer consisting of $n_w$ filters having widths $k_{w}$ and strides $dw$ are computed using~\eqref{eq:conv}.
\begin{equation}
    y_{t}^{i}=f\big(b_{i}+\sum_{k=1}^{k_{w}}w_{k}^{i} x_{{dw}\times(t-1) +k}\big)   \quad \quad  1\leq i \leq n_{w}
    \label{eq:conv}
\end{equation}
Here $f(\cdot)$ is the rectified linear function (ReLU) \cite{zeiler2013rectified}. Another important component of our classification block is a nonlinear subsampling layer. For this purpose, we use max pooling layer over time which takes the output of each convolutional layer. 
Max pooling layer reduces the temporal resolution and selects the most salient features by locally aggregating the feature map of each convolutional layer. We then concatenate the outputs of these three pooling layers to get the features with multiple temporal resolutions and provide that to the classification block.


\subsection{Classification Block}
\label{CB}

We construct our classification block by stacking CNN layer on LSTM. This is motivated by the fact that performance of LSTM can be improved by feeding it with a good representation \cite{sainath2015convolutional}. LSTM is specialised to model a long range of contexts due to their gated architecture \cite{latif2018phonocardiographic,qayyum2018quran}. Emotion in speech are context-dependent, therefore, the contexts modelling abilities of LSTM are utilised to learn the temporal structure of emotions from the given features maps. We pass the outputs of LSTM to the fully connected layers as it transforms the output of LSTM to a more discriminative space that helps the model for target prediction \cite{sainath2015convolutional}. In this way, our classification block is jointly empowered by the convolutional layer to capture high-level abstraction, the LSTM layer for long-term temporal modelling, and finally the fully connected layer for learning discriminative representations.


\section{Experimental Setup}
\subsection{Dataset}
We evaluated our model on two popular datasets: MSP-IMPROV \cite{busso2017msp} and IEMOCAP \cite{busso2008iemocap}. Both of these datasets contain dyadic interactions between actors. We only used audio recordings from these datasets. \subsubsection{IEMOCAP}
This corpus contains five sessions, where each session has utterances from two speakers (one male and one female). Overall, there are 10 unique speakers. We used four emotions including angry, happy, neutral and sad. To be consistent with previous studies \cite{aldeneh2017using}, we merged excitement with happiness and considered one class, happy.
\subsubsection{MSP-IMPROV}
The MSP-IMPROV dataset contains six sessions, where each session comprises of utterances from two speakers, one male, and one female. There are four emotion categories in MSP-IMPROV: angry, neutral, sad, happy, all were used in the experiments.

\subsection{Data Pre-processing and Augmentation}
We used the data augmentation to increase the size of training set. In particular, we created two different copies of each utterance following the approach in~\cite{ko2015audio}. For a given training utterance, we created two versions by applying the speed effect at the factors of $0.9$ and $1.1$. Sox\footnote{http://sox.sourceforge.net/} audio manipulation tool was used for data augmentation. For both datasets, we removed the non-speech intervals at the beginning and end of each utterance as was done in~\cite{ravanelli2018speech}.

\subsection{Model Configuration}
\label{Model_C}
We implemented our model using Tensorflow library. In the front end, we selected three parallel striding convolutional layers with different filter widths using the validation data. We used one layer with filter window of 25ms with a shift of 10ms to match the standard frame size of emotional feature extraction process. Smaller and larger filters (than the standard) can also extract useful information from raw speech using CNNs \cite{palaz2015analysis,muckenhirn2018towards,ghahremani2016acoustic}. Therefore, we also used two other layers with filter sizes 15ms and 100ms. The filter sizes were chosen using validation data. We used 40 filters in all three layers. We applied max pooling layer after each convolutional layer to extract the most descriptive features. The feature extraction was jointly optimised with the classification block where we used a combination of CNN and LSTM. First layer of classification block was a 2d convolutional layer, with filter size (2,2) and filter number 32, followed by the max pooling layer with the pooling size (2,2). The feature maps were then given to the LSTM layer with 128 cells for temporal modelling. Finally, we used one fully connected layer with 1024 units before the softmax layer. 

Before applying non-linearity in each convolutional layer, we used batch normalisation (BN) \cite{ioffe2015batch} layers to alleviate the problem of exploding and vanishing gradients. For regularisation, we used dropout layer after LSTM layer, with a dropout rate of 0.3. We randomly initialised the weights of our network following the techniques in ~\cite{he2015delving}. Similar to~\cite{aldeneh2017using} we trained all models using the training set, and validation set was used for hyper-parameter selection. For minimisation of cross-entropy loss function, we used RMSProp optimiser \cite{tieleman2012lecture}, with an initial learning rate of $10^{-4}$. If the UAR on the validation set did not improve after 5 epochs, we halved the learning rate. We stopped the process if the UAR did not improve for 20 consecutive epochs. For each model used in this work, we repeated the evaluation 10 times and averaged their predictions.

\section{Experiments and Results}
This section reports the experimental validation of the proposed model for speech emotion recognition. We used leave-one-speaker-out scheme for both datasets and report unweighted average recall (UAR) for both datasets. UAR is a widely used metric used for speech emotion recognition due to class imbalanced datasets. In each session, we used utterances from one speaker for testing and utterances from the other speaker for validation and early stopping~\cite{aldeneh2017using}. The remaining utterances from all speakers were used for training the model. For fair comparison with \cite{aldeneh2017using}, we used the same data augmentation technique \cite{ko2015audio} to increase the size of the training set.

For baseline results, we trained SVMs using well-known feature sets, such as, MFCC, LogMel, GeMAPS and eGeMAPS \cite{eyben2016geneva} for emotion classification. These features were extracted using openSmile toolkit \cite{eyben2013recent}. We used an RBF kernel and performed grid search using validation data to pick the optimal hyper-parameters. For a fair comparison with our model, we used the same augmented data for all SVM experiments. 
\begin{table}[!ht]
\centering
\scriptsize
\caption{UAR (\%) comparison among different models and proposed approach on raw speech}
\label{table: comp}
\begin{tabular}{lcc}
\hline
\multicolumn{1}{c}{\multirow{2}{*}{Method}} & \multicolumn{2}{c}{UAR (\%)}                                  \\ \cline{2-3} 
\multicolumn{1}{c}{}                        & \multicolumn{1}{c}{IEMOCAP} & \multicolumn{1}{c}{MSP-IMPROV} \\ \hline
SVM+MFCC   & \multicolumn{1}{l}{57.15$\pm$2.1 } & 52.38$\pm$3.7                             \\
SVM+LogMel & \multicolumn{1}{l}{58.16$\pm$2.6}& 52.54$\pm$3.1   \\
SVM+GeMAPS & \multicolumn{1}{l}{57.92$\pm$3.2}& 52.10$\pm$3.9  \\
SVM+eGeMAPS & \multicolumn{1}{l}{58.76$\pm$2.6}& 52.41$\pm$4.6 \\
CNN+MFBs \cite{aldeneh2017using} & \multicolumn{1}{l}{61.8 $\pm$3.0}& \multicolumn{1}{c}{52.6 $\pm$ 3.8 }\\
Proposed+ raw  &\multicolumn{1}{l}{60.23$\pm$3.2}& \multicolumn{1}{c}{52.43 $\pm$4.1 }\\
\hline
\hline
TDNN-LSTM+ raw (no aug) \cite{sarma2018emotion} &\multicolumn{1}{l}{48.84}   & \multicolumn{1}{c}{---}\\
SimpleNet-CNN+ raw (no aug) \cite{gong2018deep} &\multicolumn{1}{l}{52.9}  & \multicolumn{1}{c}{---}\\
Proposed+ raw (no aug)  & 56.72$\pm$3.3  & \multicolumn{1}{c}{48.54 $\pm$3.8 }\\\hline
\end{tabular}
\end{table}

Table \ref{table: comp} shows the comparison of results using different methods, and also shows the comparison with previous studies \cite{aldeneh2017using,sarma2018emotion,gong2018deep}. A direct comparison with some of these studies is not possible due to the difference in data augmentation methods used in the studies, which may affect the results. For example, \cite{sarma2018emotion} used different data augmentation scheme, while Gong et al. \cite{gong2018deep} did not use any data augmentation. We therefore compare our results with these studies \cite{aldeneh2017using,sarma2018emotion,gong2018deep} without any data augmentation and separate from other results using a double line in Table \ref{table: comp}. 

\section{Analysis and Discussion}

\subsection{Convolutional Layers Analysis}
The convolutional layers play a crucial role in the performance of emotion recognition from raw speech \cite{sarma2018emotion,ravanelli2018speech}. It is interesting to see the effect of using parallel convolutional layers for capturing multi-temporal resolution features from the raw speech in the feature extraction block. We evaluated different number (1,2,3,4) of parallel convolutional layers and reported the associated results in Table \ref{table: layers}. 
\begin{table}[!ht]
\centering
\scriptsize
\caption{Effect of using different number of parallel convolutional layers}
\label{table: layers}
\begin{tabular}{ccc}
\hline
\multicolumn{1}{c}{\multirow{2}{*}{Layers}} & \multicolumn{2}{c}{UAR (\%)}                                  \\ \cline{2-3} 
\multicolumn{1}{c}{}                        & \multicolumn{1}{c}{IEMOCAP} & \multicolumn{1}{c}{MSP-IMPROV} \\ \hline
  1 &57.36$\pm$2.3&48.36$\pm$3.1 \\
 2 &58.32$\pm$2.8&50.12$\pm$3.5 \\
  3 &60.23$\pm$3.2 &52.43$\pm$4.1 \\
  4 &59.13$\pm$3.1&52.21$\pm$4.0\\\hline
\end{tabular}
\end{table}
The results show that the proposed multi-temporal resolution model with parallel convolutional layers outperforms the single layer architecture. The best results are obtained using 3 parallel layers, which suggests that a suitable number of parallel layers needs to be determined empirically for specific problems.


\subsection{Pooling Strategies}
Since the pooling layer is used for generalisation of time-domain averaging \cite{sainath2015learning}, we evaluated three different pooling operations including max, $l_{2}$ and average. Table \ref{table: pooling} shows that max pooling outperformed others. All results reported in this paper, therefore, use max pooling. 
\begin{table}[!ht]
\centering
\scriptsize
\caption{UAR (\%) with different pooling strategies.}
\label{table: pooling}
\begin{tabular}{lcc}
\hline
\multicolumn{1}{c}{\multirow{2}{*}{Pooling}} & \multicolumn{2}{c}{UAR (\%)}                                  \\ \cline{2-3} 
\multicolumn{1}{c}{}                        & \multicolumn{1}{c}{IEMOCAP} & \multicolumn{1}{c}{MSP-IMPROV} \\ \hline
max  & 60.23$\pm$3.2& \multicolumn{1}{c}{ 52.43$\pm$4.1 }   \\
$l_{2}$ & \multicolumn{1}{c}{59.72$\pm$2.8}         &     50.25$\pm$3.0    \\
Average              & \multicolumn{1}{c}{ 59.50$\pm$3.0 }        & 51.94 $\pm$3.2\\\hline                        
\end{tabular}
\end{table}
\subsection{Analysing Classification Block}
In this section, we analyse the effect of using different type of layers in the classification block. Results are reported in Table \ref{layers} for both datasets using different configuration of layers in the classification block. In all these setups, we use the same feature extraction block consisting of three parallel convolutional layers that provide multiple temporal dependencies. The architectural changes are only made on the classification block.  We trained different configuration of classification blocks including three DNNs (1024-512-512), two LSTMs (256-256), and three CNN layers (256 feature map). We also evaluated other combinations, such as, LSTM-DNN (2 LSTM, 1 DNN), CNN-DNN (2 CNN, 1 DNN), CNN-LSTM (1 CNN, 2 LSTM), and CNN-LSTM-DNN (1 CNN, 1 LSTM, and 1 DNN). All these combinations were trained using the evaluation recipe described in Section \ref{Model_C}.  
\begin{table}[!ht]
\centering
\scriptsize
\caption{Analysis of results (UAR) using different combination of layers in classification block}
\label{layers}
\begin{tabular}{lcc}
\hline
\multicolumn{1}{l}{\multirow{2}{*}{Method}} & \multicolumn{2}{c}{UAR (\%)}                                  \\ \cline{2-3} 
\multicolumn{1}{c}{}                        & \multicolumn{1}{c}{IEMOCAP} & \multicolumn{1}{c}{MSP-IMPROV} \\ \hline
 DNN &53.36$\pm$2.0&48.36$\pm$3.2 \\
 LSTM-DNN &56.32$\pm$2.6&49.58$\pm$3.0 \\
 LSTM&58.72$\pm$2.9&51.21$\pm$3.4 \\
 CNN-DNN &58.43$\pm$2.8&50.44$\pm$3.1 \\
 CNN-LSTM&59.23$\pm$3.0&52.36$\pm$3.6 \\
 CNN-LSTM-DNN&\textbf{60.23$\pm$3.2}&\textbf{52.43$\pm$4.1}\\
 CNN&58.52$\pm$2.6&50.84$\pm$3.6 \\\hline
\end{tabular}
\end{table}

In Table \ref{layers}, we observe that only using DNN layers in classification block hurts the performance of the model. However, their combination with LSTM and CNN is beneficial for the model performance. We achieved the best performance with the classification block while using the combination of CNN, LSTM and DNN layer. This shows that our proposed construct of the classification block, where convolutional layer captures high-level abstraction, LSTM layer performs long-term temporal modelling, and the fully connected layer performs discriminative representations, offers improvements in the performance of emotion recognition.  
\subsection{Input Length Analysis}
\begin{figure}[!ht]
  \centering
  \includegraphics[width=0.8\linewidth]{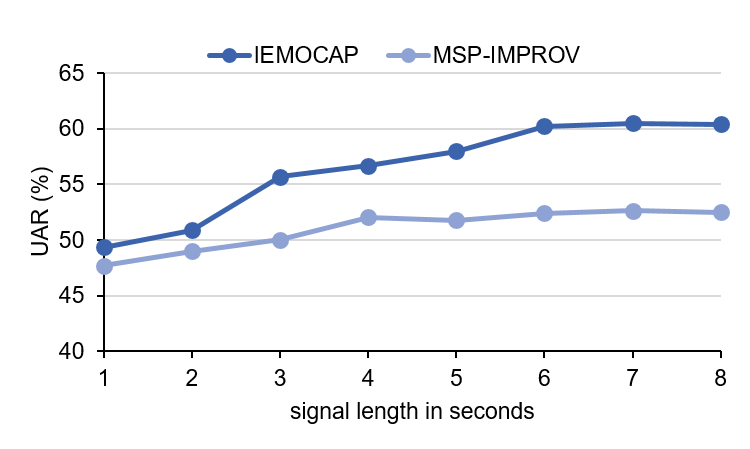}
  \caption{System performance with different signal length on both datasets.}
  \label{fig:signal}
\end{figure}

We evaluated the performance of our proposed model for different signal lengths. Results are presented in Figure \ref{fig:signal}. 
Signal length is an important aspect since short signals create many small segments which need to be merged, whereas  long signals can potentially cause buffer overflow in embedded systems with limited memory. For both IEMOCAP and MSP-IMPROV, UAR increases with the increase in speech signal length in general. When the signal length is small (1 or 2 seconds) emotion recognition can be performed with a small accuracy loss. 
However, we observe that speech utterance signal of 6 seconds offers the best UAR for both datasets.

\subsection{Classification Performance Analysis}
We compare the results of our proposed model with that of SVM trained on widely used state-of-art feature sets including MFCC, LogMel, GeMAPS, eGeMAPS in Table \ref{table: comp}. It can be observed that our proposed architecture modelling multi-temporal feature from raw speech can achieve better performance compared to the powerful classifier SVM using state-of-the-art features.   

We also compare our results with three relevant studies~\cite{aldeneh2017using,sarma2018emotion,gong2018deep} and present the results in Table \ref{table: comp}. In \cite{aldeneh2017using} authors used Mel Filterbank (MFB) features as the input to CNNs and showed that CNNs with these hand-engineered features can produce competitive results to the popular feature sets. In contrast, we used raw speech as input to the model and jointly optimised the feature extraction with the classification network. We achieved comparable results with this study on both datasets as presented in Table \ref{table: comp}. This shows that capturing multi-temporal dependencies from the raw speech using parallel CNN layers helps to achieve comparable performance to CNNs trained on hand-engineered features. Two other recent studies \cite{sarma2018emotion,gong2018deep} used raw speech and evaluated their approach on IEMOCAP dataset. We are achieving better results compared to them when data augmentation is not used.  Other recent studies \cite{yenigalla2018speech,etienne2018cnnlstm} used CNN based models and achieved UAR of  61.9\% and 61.7\% on IEMOCAP dataset using spectrograms as the input. Compared to these studies, we are achieving 60.23\% directly using raw speech.

\section{Conclusions}
In this paper, using two widely used emotion corpus: IEMOCAP and MSP-IMPROV, we show that the proposed network of parallel multi-layer CNN stacked on an LSTM offers, (1) better accuracy when compared to existing methods using raw speech waveform for emotion recognition and (2)  comparable accuracy to existing methods using state-of-the-art hand-engineered features. We claim that our proposed construct of CNN having parallel convolutional layers with multiple filter lengths capture both long-term and short-term interactions and help us achieve this performance. In our future studies, we aim to further investigate ways to improve emotion recognition accuracy using raw speech. We also aim to perform run-time and computational complexity comparisons between methods using raw-speech and hand-engineered features and report the accuracy-complexity trade-off.







\end{document}